\documentclass[useAMS,usenatbib]{mn2e}
\usepackage{graphics}
\usepackage{graphicx}
\usepackage{epsf}
\usepackage{url}

\newcommand{\bk}{{\bmath k}}

\def\hMpc{{\ }h^{-1}{\,}{\rm Mpc}}
\def\ihMpc{{\ }h{\,}{\rm Mpc}^{-1}}
\newcommand{\expect}[1]{\left\langle #1 \right\rangle}  % expectation value

\newcommand{\infomassfn}{
  \begin{figure}
    \begin{center}
     \leavevmode
      \epsfxsize=\columnwidth
      \epsfbox{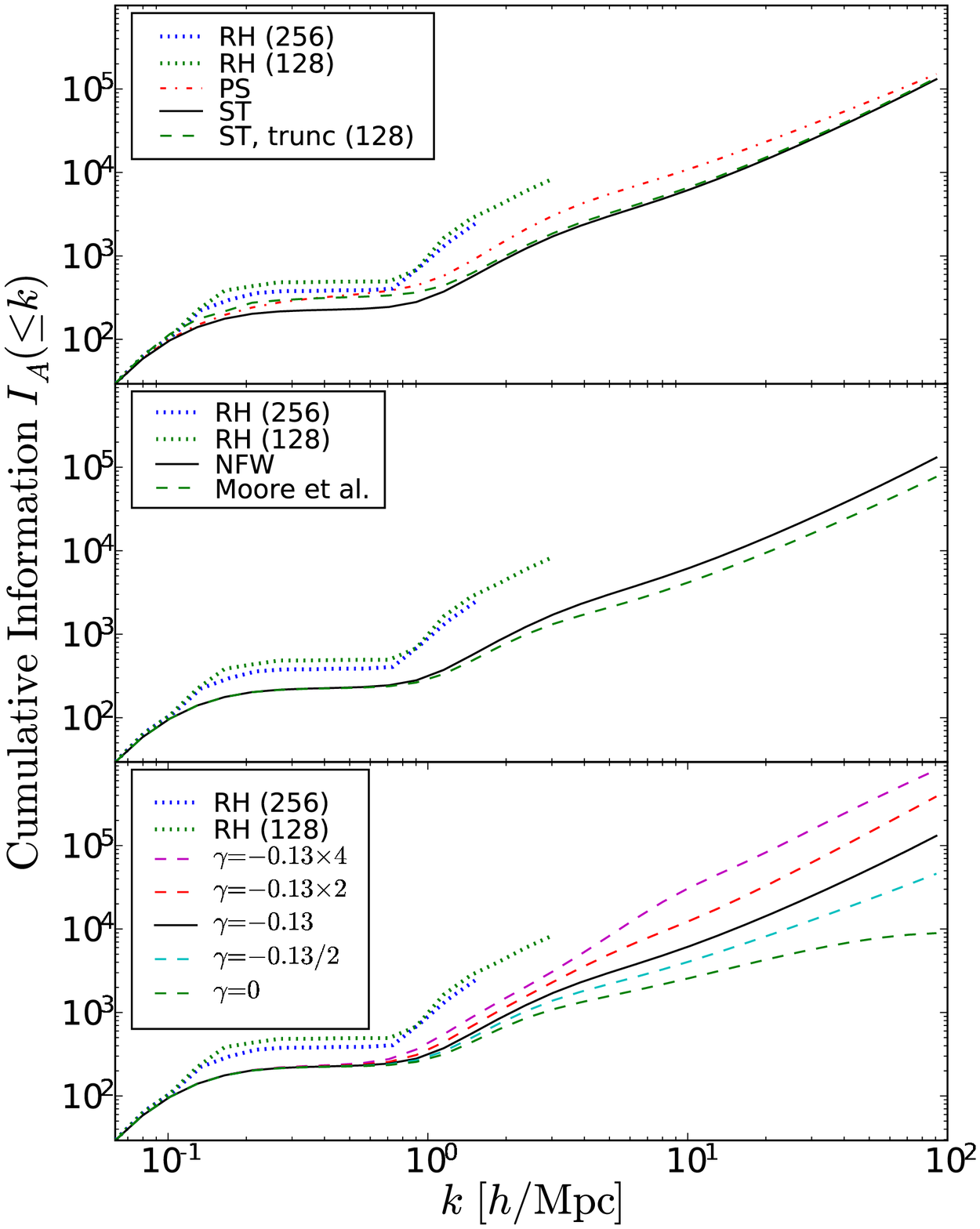}
    \end{center}

    \caption[1]{ \small The Fisher information in the halo-model
    dark-matter power spectrum up to wavenumber $k$, as defined in
    Eq.\ \ref{ibk}, for different variants of the halo model.  In each
    panel, the dotted curves show what RH found from hundreds of
    $N$-body simulations of box sizes 128 and 256$\hMpc$.  (For
    consistency, for the $128\hMpc$ box size, we use only the results
    of their 200 particle-mesh simulations.)  Like RH, we normalize
    all $I_A(\leq k)$ for a $256\hMpc$ box.  The solid black lines are
    from the fiducial model, with all halo-model inputs the same as in
    CH except for the mass function, which we take from \citet{st}; CH
    used a \citet{ps} mass function.  The top panel shows that the
    mass function controls the shape of the translinear plateau.  The
    dashed green curve is from an ST mass function calculated using
    $P^{\rm lin}(k < \frac{2\pi}{128 \hMpc}) = 0$, truncated to mimic
    the conditions in a finite box.  Changes in the small-scale slope
    of halo density profiles (middle panel) or in the slope $\gamma$
    of the relationship between halo mass and concentration parameter
    (bottom panel) alter the small-scale shape of $I_A(\leq k)$.  In
    the bottom panel, the curves rise as $\gamma$ grows more
    negative.

    \label{infomassfn} 

	}
   \end{figure}
}

\newcommand{\infozphys}{
  \begin{figure}
    \begin{center}
     \leavevmode
      \epsfxsize=\columnwidth
      \epsfbox{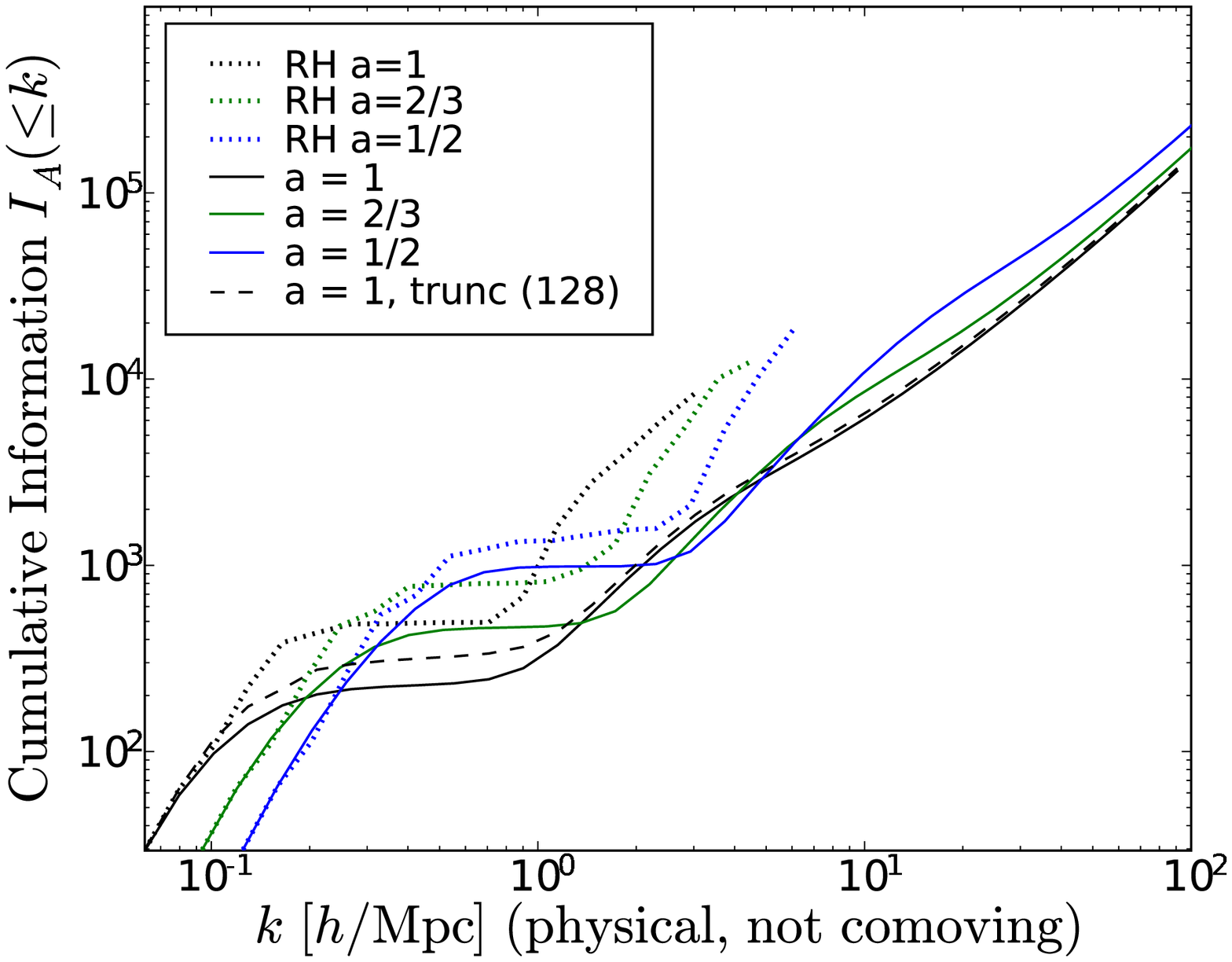}
    \end{center}

    \caption[1]{ \small The information content of the halo-model
    power spectrum about the initial amplitude of the linear power
    spectrum, as a function of physical scale, for various redshifts.
    Also shown are the results from 200 particle-mesh simulations by
    RH, for boxes of size $128\hMpc$.  For $a=1$, we also show the
    result if the linear power spectrum used for the ST mass
    function is truncated to mimic a box of this size.  The
    information up to the physical scale of $100\ihMpc$ in this
    fiducial model decreases slowly with time.
\label{infozphys}
    }
  \end{figure} }

\newcommand{\pleisiosaur}{
  \begin{figure}
    \begin{center}
     \leavevmode
      \epsfxsize=\columnwidth
      \epsfbox{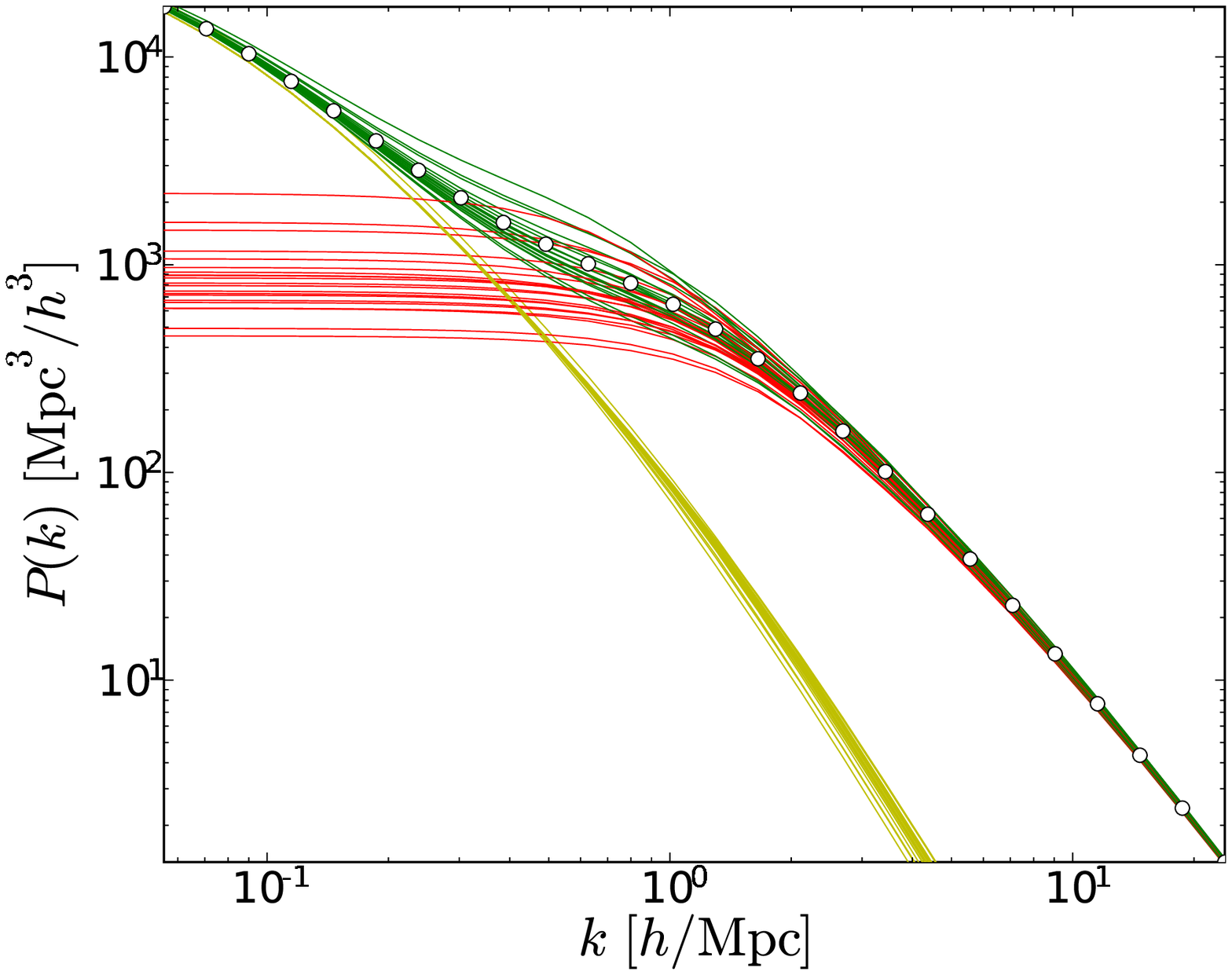}
    \end{center}

    \caption[1]{ \small Power spectra from twenty realizations of an
    ST halo mass function for a volume $V = (128\hMpc)^3$.  The number
    of haloes in each mass bin $\delta(m)$ is drawn from a Poisson
    distribution of mean $Vn(m)\delta(m)$.  The fluctuated 1- and
    2-halo terms, and their sum, are shown in red, yellow, and green,
    respectively; the unfluctuated power spectrum appears as white
    dots.  (The 2-halo terms are renormalized to agree with the linear
    power spectrum in the largest-scale bin.)  There is a great deal
    of variance, and covariance, in the translinear regime, since it
    is here that Poisson fluctuations in the largest (rarest) haloes
    have the greatest influence on the power spectrum.
\label{pleisiosaur} }
\end{figure} }

\newcommand{\infoterms}{
  \begin{figure}
    \begin{center}
     \leavevmode
      \epsfxsize=\columnwidth
      \epsfbox{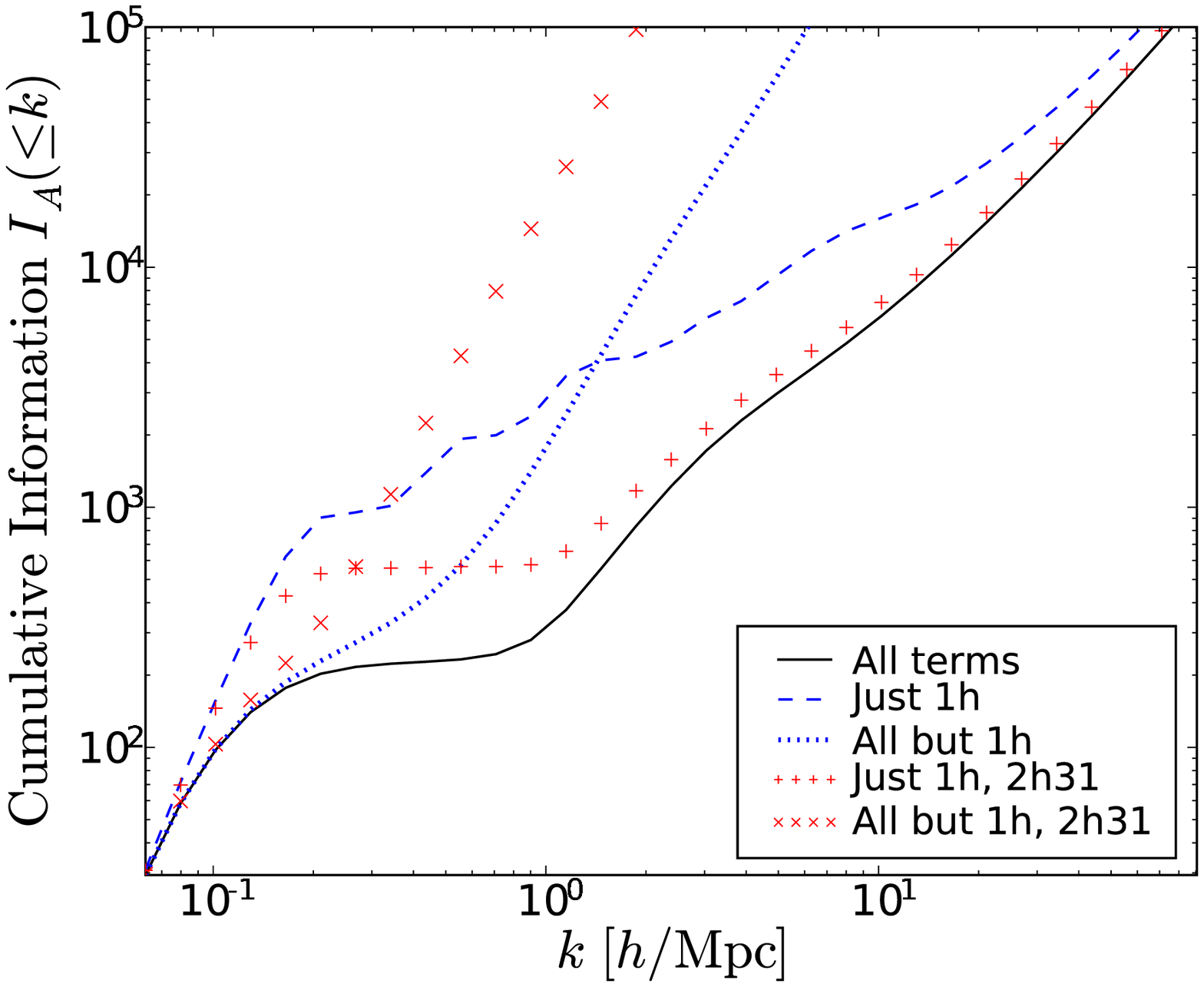}
    \end{center}

    \caption[1]{ \small The information content in the halo-model
    power spectrum about the initial amplitude of linear power, if
    only $T^{\rm 1h}$, and $T^{\rm 2h}_{31}$ along with $T^{\rm 1h}$,
    are included in the trispectrum.  Also shown is the information
    content if all halo-model trispectrum terms except these are
    included.  The Gaussian piece of the covariance, using the full
    halo-model power spectrum, is always included.  When covariance
    matrix terms are added, $I_A(\leq k)$ decreases, but not simply
    additively, since the Fisher matrix is a matrix inverse of a sum,
    not a sum of matrix inverses.  The 1h term can be attributed to
    Poisson fluctuations in the halo mass function, as can, perhaps,
    $T^{\rm 2h}_{31}$.  These two terms dominate the behavior of the
    information curve, indicating that Poisson halo fluctuations may
    largely be responsible for the loss of information in the
    non-linear regime.

    \label{infoterms} }
    \end{figure} }

\begin{document}
\title[Halo-model dark-matter power spectrum information content]
{Information content in the halo-model dark-matter power spectrum}

\author[Mark C.\ Neyrinck, Istv\'{a}n Szapudi and Christopher D.\ Rimes]
{Mark C.\ Neyrinck$^1$, Istv\'{a}n Szapudi$^1$ and Christopher D. Rimes$^2$\\
$^{1}$IfA, University of Hawaii, Honolulu, HI 96822, USA\\
$^{2}$JILA, University of Colorado, Boulder, CO 80309, USA\\
  email: {\tt neyrinck@ifa.hawaii.edu}}
\pubyear{2006}

\bibliographystyle{mnras}
	 
\maketitle
  
\begin{abstract}
  Using the halo model, we investigate the cosmological Fisher
  information in the non-linear dark-matter power spectrum about the
  initial amplitude of linear power.  We find that there is little
  information on `translinear' scales (where the one- and two-halo
  terms are both significant) beyond what is on linear scales, but
  that additional information is present on small scales, where the
  one-halo term dominates.  This behavior agrees with the surprising
  results that \citet{rh05,rh06} found using $N$-body simulations.  We
  argue that the translinear plateau in cumulative information arises
  largely from fluctuations in the numbers of large haloes in a finite
  volume.  This implies that more information could be extracted on
  non-linear scales if the masses of the largest haloes in a survey
  are known.
\end{abstract}

\begin {keywords}
  cosmology: theory -- large-scale structure of Universe -- dark matter
\end {keywords}

\section{Introduction}
Valuable cosmological information is encoded in the large-scale
structure of the Universe.  Rimes \& Hamilton (2005; 2006, RH) have
studied how much cosmological Fisher information is in the non-linear
dark-matter power spectrum.  As a start, they considered the
information about the initial amplitude $A$ of linear power.  RH
measured this information from the covariance matrix of power spectra
from $N$-body simulations.  \citet{mw}, \citet[][SZH]{szh} and
\citet[][CH]{chu} have also explored this covariance matrix, without
explicitly using the information concept.  RH found that, as expected,
information is preserved on large, linear scales, but the behavior is
surprising on smaller scales.  On translinear scales ($k \sim
0.2-0.8\hMpc$), the information is mostly degenerate with that in the
linear regime, but there is significant extra information on smaller
scales.  This does not mean that surveys reaching only translinear
scales are useless, but that translinear scales contribute little
extra information about $A$ in a survey extending into the linear
regime.  Such clearly demarcated behavior in different regimes is
suggestive of the halo model, in which the power spectrum consists of
a term on linear scales and a term from virialized haloes.
 
\section{Method}
The Fisher information about a parameter $\alpha$ given a set of data
is defined \citep[e.g.][]{tth} as
\begin{equation}
  I_\alpha \equiv - \expect{ \frac{\partial^2\ln\mathcal{L}(\alpha|{\rm data})}{\partial \alpha^2}},
\label{iabstract}
\end{equation}
where $\mathcal{L}$ is the likelihood function of $\alpha$ from the
data.

We discuss the information in the (hereafter, implicitly non-linear
dark-matter) power spectrum $P(k)$ about the logarithm of the initial
amplitude $A$ of the linear power spectrum $P^{\rm lin}(k)$.  In this
paper, the word `information' is implicitly about $A$, although this
framework also allows other parameters to be investigated.  The
information in measurements $P(k_i)$ and $P(k_j)$ of the power
spectrum is (RH)
\begin{equation}
  I_A = - \expect{ \frac{\partial\ln P(k_i)}{\partial\ln A}
  \frac{\partial^2\ln\mathcal{L}}{\partial\ln P(k_i) \, \partial\ln
  P(k_j)} \frac{\partial\ln P(k_j)}{\partial\ln A} }.
\label{ipkikj}
\end{equation}
We approximate the expectation value of the middle term on the right
side of this equation (i.e.\ the Fisher matrix) with the inverse of
the covariance matrix, $[C^{-1}]_{ij}$. Here, $C_{ij} \equiv
\expect{\Delta P(k_i)\Delta P(k_j)}$, where $\Delta P(k_i)$ is a
fluctuation of an estimate away from the mean in $P(k_i)$.  This
approximation is good if the distribution of estimates of power about
their mean is Gaussian, which RH showed to be sufficiently so in this
case.

We follow the procedure of RH to measure $I_A$.  We use uncorrelated
band powers $B(k_i)$ with a diagonal Fisher matrix.  The band powers
are decorrelated in such a way that they retain the same expectation
values as $P(k_i)$.  Following RH, we use an upper-Cholesky
decomposition, ensuring that the information up to a given $k_{\rm
max}$ is affected only by data for $k\leq k_{\rm max}$.  The
cumulative information $I_A(\leq k)$ which we measure in terms of band
powers is then
\begin{equation}
  I_A(\leq k) = - \expect{ \sum_{k=0}^{k_{\rm max}} \frac{\partial\ln
      B(k)}{\partial\ln A} \frac{\partial^2\ln\mathcal{L}}{\partial\ln
      B(k)^2} \frac{\partial\ln B(k)}{\partial\ln A}}.
\label{ibk}
\end{equation}

\subsection{Covariance matrix construction}
The covariance of the power spectrum in a survey of volume $V$ is the
sum of a Gaussian term, which depends on the square of the power
spectrum itself, and a term involving the (hereafter, implicitly
non-linear) trispectrum \citep[SZH;][]{hrs};

\begin{equation}
C_{ij} = \frac{1}{V}\left[\frac{(2\pi)^3}{V_{s,i}}2P(k_i)^2\delta_{ij}
+ T_{ij}\right],
\label{cijdef}
\end{equation}
where $V_{s,i}$ is the volume of shell $i$ in Fourier space
(proportional to $k_i^3$ for logarithmically spaced bins), and $T_{ij}$
is the trispectrum averaged over shells $i$ and $j$;
\begin{equation}
T_{ij} \equiv \int_{s,i}\int_{s,j}T(\bk_i,-\bk_i,\bk_j,-\bk_j)\frac{d^3 \bk_i}{V_{s,i}}\frac{d^3\bk_j}{V_{s,j}}.
\label{angav}
\end{equation}

To obtain the power spectrum and trispectrum, we use the halo-model
trispectrum formalism developed by CH, which we sketch here.  In the
halo model, the matter power spectrum is the sum of one- and two-halo
terms.
\begin{eqnarray}
P(k) & = & P^{\rm 1h}(k) + P^{\rm 2h}(k)\\
     & = & M^0_2(k,k) + P^{\rm lin}(k)[M^1_1(k)]^2,
\label{p1h2h}
\end{eqnarray}
where $M^\beta_\mu$ are integrals over the halo distribution;
\begin{eqnarray}
M^\beta_\mu(k_1,\ldots,k_\mu) & \equiv & \int\int \left(\frac{m}{\bar{\rho}}\right)^\mu b_\beta(m)n(m,c)\nonumber\\
& &\times u(k_1,m,c)\cdots u(k_\mu,m,c)\,dc\,dm.
\label{ibetamu}
\end{eqnarray}
Here, $\bar{\rho}$ is the mean matter density, $m$ is the halo mass,
$c$ is the halo concentration as in the NFW profile \citep{nfw96},
$b_\beta(m)$ is the $\beta$-order halo bias \citep{mjw,sshj}, and
$u(k,m,c)$ is the halo profile in Fourier space, normalized to unity
at $k=0$.  Like CH, we use a distribution in halo concentration since
its effect can be large for higher-order correlations.

The halo-model trispectrum is the sum of {one-,} {two-,} {three-,} and
four-halo terms \citep[CH;][appendix]{coorayphd}.  We quote those
terms which seem to have a simple physical explanation (Sect.\ \ref{phaf}): the one-halo term,
\begin{equation}
T^{\rm 1h}(\bk_i,-\bk_i,\bk_j,-\bk_i) = M^0_4(k_i,k_i,k_j,k_j)
\label{t1h}
\end{equation}
(the arguments of $T$ terms will be implicit henceforth), and the
component of the two-halo term which comes from taking three points in
one halo and one point in the second,
\begin{equation}
T^{\rm 2h}_{31} = P^{\rm lin}(k_i)M^1_3(k_i,k_j,k_j)M^1_1(k_i) + 3{\rm\ perm}.
\label{t2h31}
\end{equation}

In implementing the halo-model trispectrum, we use the same
cosmological parameters as RH: $(\Omega_{\rm M},\Omega_\Lambda, f_{\rm
b}, h, \sigma_8) = (0.29,0.71,0.16,0.71,0.97)$.  The only halo-model
parameter change from CH in our fiducial model is our use of the
\citet[][ST]{st}, instead of the \citet[]{ps}, mass function.  We
evaluate the trispectrum in the center of each bin, performing only an
angular average.  For the fiducial model, we tried decreasing the bin
size by a factor of five, which barely affected the cumulative
information.

\infomassfn

Using all the same parameters as CH, we obtained excellent agreement
with the power spectrum they plotted.  We also verified that our code
gives the correct relevant $M^\beta_\mu$ functions in an analytical
version of the halo model \citep[][$\S4.3$]{cs}.  We used the
convenient form of the angular-averaged perturbation-theory
trispectrum provided by SZH (their eq.\ 7), and checked our
perturbation-theory trispectrum against simple results they gave.
Also, our $T^{\rm 1h}$ agrees with the covariance in $P^{\rm 1h}$ due
to halo Poisson sampling, as theoretically expected (see appendix).
However, our square-configuration trispectrum $\Delta^2_{\rm sq}(k)$
was slightly lower than that plotted by CH; also, our correlation
matrix $\frac{C_{ij}}{\sqrt{C_{ii}C_{jj}}}$ appears to reach a given
value at a slightly larger $k$ than in Table 1 of CH, by a factor of
$\sim 1.5$.  But these discrepancies are noticeable in the regime
where the simple, well-tested $T^{\rm 1h}$ term dominates.  After
extensive tests, we are confident that our code is accurate.  In any
case, our qualitative result is insensitive to these slight
differences.

\section{Results}

Figure \ref{infomassfn} shows the cumulative information $I_A(\leq k)$
as found with the halo model under different sets of parameters, and
as measured by RH from $N$-body simulations.  All curves share these
qualitative features: on linear scales, $I_A(\leq k) \propto k^3$, as
expected; on translinear scales, there is little additional
information; and on the smallest scales, the information rises again,
but less steeply than in the linear regime.  This small-scale rise
seems to be shallower in the halo model than in simulations, although
the small-scale measurements by RH were near their resolution limit.

Generally, the halo mass function affects the details of the
translinear plateau in the cumulative information $I_A(\leq k)$, and
halo density profiles affect the small-scale upturn.  Changing the
inner slope of the halo density profile between $-1$ (NFW) and $-1.5$
\citep{moore}, fixing the density at the scale radius $r_s$, gives
little change in $I_A(\leq k)$.  We changed the slope $\gamma$ in the
relation $c(m) = c_0 (m/m_\star)^\gamma$ between halo mass and
concentration parameter more drastically; varying $\gamma$ from zero
to four times its fiducial value, $-0.13$ \citep{bull}, has a large
effect on $I_A(\leq k)$ on small scales.  When $c(m)$ does not change
with $m$, there is an apparent plateau in $I_A(\leq k)$ at the
smallest scales, while $I_A(\leq k)$ is greatest when $c(m)$ falls off
most sharply.  Therefore, it could be said that the information on
small scales is encoded in the systematic variation in halo density
profiles with halo mass.

\infozphys

An interesting question which RH posed is whether the information
(e.g., about $A$) in the power spectrum is preserved in time.  If it
is, then the cumulative information measured up to a fixed large
physical (not comoving) wavenumber should remain constant in time.  It
would not make sense if information were created, but non-linear
evolution could decrease with time; for example, it could divert into
higher-order statistics.  Requiring information to be preserved or
lost is a test of the model which could be used to constrain
halo-model parameters.

Our fiducial model passes this test; Figure \ref{infozphys} shows
information curves as a function of physical $k$ for the fiducial
model at fairly low redshifts (where the halo model is well-tested),
along with curves from RH for comparison.  Looking at the right edge
of the figure, it seems that the halo model predicts that information
up to a fixed physical scale of $100\ihMpc$ gradually decreases with
time.

\subsection{Physical explanation: fluctuations in the number of large haloes in a finite volume}\label{phaf}

One reason to investigate the information concept with an analytic
model is to gain physical insight.  Much of the behavior of the
cumulative information seems to come from the following Poisson halo
abundance fluctuation (PHAF) effect: in a finite volume $V$, the
number of haloes of mass between $m$ and $m+dm$ is not $Vn(m)dm$
(which would give fractional haloes), but is an integer drawn from a
distribution with mean $Vn(m)dm$.  We assume that the distribution is
adequately Poissonian for haloes large enough that only a few may
occur in the volume, and that the large number of smaller haloes is
stable enough that non-Poissonianity in the distribution has a
negligible effect.

In fact, the one-halo term $T^{\rm 1h}$ of the halo-model covariance
corresponds to the covariance $\expect{\Delta P^{\rm 1h}(k_i)\Delta
P^{\rm 1h}(k_j)}$ in $P^{\rm 1h}$ from Poisson-sampling the number of
haloes in each bin.  This can be shown analytically by considering
mass bins small enough to contain only zero or one halo (see
appendix).  We also verified this numerically by comparing $T^{\rm
1h}$ to the covariance of $P^{\rm 1h}$ for 262144 Poisson realizations
of the mass function.  Compared to the true CH covariance in eq.\
(\ref{cijdef}), we consider the PHAF covariance to be an approximation
which is not entirely accurate, but which shares two terms (see
appendix) with the true covariance.

\pleisiosaur

Figure \ref{pleisiosaur} shows the power spectra from twenty Poisson
realizations of the ST mass function in a volume $V = (128\hMpc)^3$.
(Fluctuating the mass function also causes $P^{\rm 2h}$ to fluctuate
on large scales, but we have renormalized $P^{\rm 2h}$ to match
$P^{\rm lin}$ in the largest-scale bin.)  This pleisiosaur shape helps
explain the shape of the information curve intuitively.  Fluctuations
in the number of the largest, rarest haloes in a finite volume lead to
large variances (and covariances) in $P^{\rm 1h}$ on large scales.  On
translinear scales, this effect squelches the information.  On larger,
linear scales, information is preserved since $P^{\rm 2h}$ dominates,
washing out the fluctuations in $P_{\rm 1h}$.  On fully non-linear
scales, the power spectrum is dominated by power from small haloes,
which are more stable in their numbers, giving less covariance, and
therefore significant information.

This pleisiosaur shape can also be understood as a dispersion in where
$P(k)$ turns away from $P^{\rm lin}(k)$.  This suggests that the HKLM
`scaling {\it Ansatz}' \citep{hklm,pd} may still be valid
conceptually, despite RH's apparent evidence to the contrary, if the
dispersion in the function taking linear to non-linear power is
properly taken into account.

Figure \ref{infoterms} shows how large a role $T^{\rm 1h}$ plays in
the halo-model covariance.  Keeping the Gaussian part from both terms
of $P(k)$ fixed, but using only $T^{\rm 1h}$ for $T$, the information
$I_A^{\rm 1h}(\leq k)$ shares some features with the full $I_A(\leq
k)$.  On small scales, $T^{\rm 1h}$ dominates $T$, and it is still
significant on translinear scales.  But this is not the whole story,
since the translinear plateau of $I_A^{\rm 1h}(\leq k)$ is indistinct
and wiggly.  We have not determined the source of the wiggles, but odd
behavior in $I_A^{\rm 1h}(\leq k)$ is not necessarily worrisome, since
it uses only part of the physical $T$, yet uses the Gaussian
covariance from the full $P(k)$.

\infoterms

There is also a term $\expect{P^{\rm 1h}(k_i)P^{\rm 2h}(k_j)}$ in the
halo abundance fluctuation covariance (see appendix) which equals
$T^{\rm 2h}_{31}$ (eq.\ \ref{t2h31}).  However, it is less certain
that $T^{\rm 2h}_{31}$ may be attributed to PHAF.  We measured the
PHAF covariance of the full power spectrum, and found off-diagonal
correlations near unity in the linear regime, which do not occur in
linear theory.  These high correlations come from $\expect{P^{\rm
2h}(k_i)P^{\rm 2h}(k_j)}$, in which mass function fluctuations cause
$P^{\rm 2h}$ to be biased relative to $P^{\rm lin}$ by a constant (in
$k$) on large scales.  (This bias was removed in Fig.\
\ref{pleisiosaur}.)  The error likely comes from replacing the bias
$b_2(m)$ in the (true) CH covariance with $[b_1(k)]^2$ in the PHAF
covariance.  This replacement does not occur in $T^{\rm 2h}_{31}$, so
we still attribute $T^{\rm 2h}_{31}$ to PHAF, but less confidently
than we do $T^{\rm 1h}$.  Figure \ref{infoterms} shows what happens if
$T^{\rm 2h}_{31}$ is included or excluded along with $T^{\rm 1h}$;
these two terms dominate the behavior of $I_A(\leq k)$.

\section{Discussion}
The qualitative agreement we found between power spectrum information
content (about the initial amplitude $A$ of linear power) in the halo
model and in $N$-body simulations (RH) is a success for the halo model
paradigm describing the power spectrum and trispectrum.  Our results
also lend weight to the argument made by RH that the small-scale
upturn in cumulative information, seen in their simulations, has a
real, physical origin and is not the result of insufficient
resolution.  However, the agreement is not perfect, as with the
bispectrum \citep{fps}, so more halo-model ingredients, such as halo
triaxiality \citep[e.g.][]{sws}, may be necessary for the halo model
to reach high precision for higher-order statistics.  On the other
hand, the high (co)variance on translinear scales suggests that
inaccuracies in the halo model on these scales have less statistical
significance than they seem.

Even if our information curves do not match exactly what RH found, the
halo model allows us to understand their results better.  The halo
model will also allow the information about other cosmological
parameters than $A$ to be studied easily.  With a covariance matrix
such as we used here, all that is needed to investigate another
parameter $\alpha$ is $\frac{\partial P(k_i)}{\partial \alpha}$,
which the halo model can provide self-consistently.

Fluctuations in the number of haloes of a given mass (particularly,
massive and rare ones) in a finite volume appear to be largely
responsible for the paucity of cumulative information on translinear
scales.  Therefore, prospects for extracting cosmological information
from the power spectrum on these scales might improve substantially
with prior knowledge of the mass spectrum in a survey.  For example, a
conditional power spectrum depending on the mass of the largest
cluster in a survey could contain significantly more information on
non-linear scales than the power spectrum alone.

\section*{Acknowledgments}
We thank Andrew Hamilton, Wayne Hu and Nick Gnedin for helpful
discussions, and an anonymous referee for suggestions.  We are
supported by NASA grants AISR NAG5-11996, ATP NAG5-12101 (MCN, IS) and
ATP NAG5-10763 (CDR), and NSF grants AST-0206243, AST-0434413, ITR
1120201-128440 (MCN, IS) and AST-0205981 (CDR).

\appendix
\section{Covariance from halo mass function fluctuations}

Often in the halo model, integrals over halo mass functions are
multiplied together.  Formally, Poisson-sampling the halo mass
function introduces an extra term in the expectation value of these
products.  Consider the ensemble average
\begin{equation}
\expect{\int f(m)\tilde{n}(m)\,dm\int g(m)\tilde{n}(m)\,dm}
\label{fgint}
\end{equation}
over fluctuations $\tilde{n}(m)$ of the mean halo number density $n(m)$,
where $f(m)$ and $g(m)$ are functions (e.g.\ halo density profiles) of
mass.  A Poisson realization of the mass spectrum may be considered,
in a way reminiscent of Peebles' (1980) derivation of power spectrum
shot noise, by partitioning the mass function into bins $\delta(m_i)$
small enough that in a given volume $V$, the number of haloes in each
bin, $\phi_i \equiv V\tilde{n}(m_i)\delta(m_i)$, is 0 or 1.
Expression (\ref{fgint}) becomes

\begin{eqnarray}
\frac{1}{V^2}\expect{\sum_{i,j} f_ig_j\phi_i\phi_j} = \frac{1}{V^2}\expect{\sum_{i\not= j}f_ig_j\phi_i\phi_j}\nonumber\\
+ \frac{1}{V^2}\expect{\sum_i f_i g_i \phi_i},
\label{fgphi}
\end{eqnarray}
since $\phi_i^2 = \phi_i$.  Returning to integral notation, this is
\begin{equation}
\int f\,n\,dm\int g\,n\,dm + \frac{1}{V}\int fg\,n\,dm.
\label{ifigifg}
\end{equation}

Products of integrals over the mass function abound in the covariance
$\expect{\Delta P_i\Delta P_j}$ of the halo-model power spectrum (eq.\
\ref{p1h2h}), where $\Delta P(k) = \tilde{P}(k) - P(k)$.  Considering
the product of two fluctuations in $P^{\rm 1h}$,
\begin{equation}
\expect{\Delta P^{\rm 1h}(k_i)\Delta P^{\rm 1h}(k_j)} = \frac{1}{V}M^0_4(k_i,k_i,k_j,k_j),
\label{p1hp1h}
\end{equation}
which happens to be the one-halo term of the full CH covariance, eq.\
(\ref{t1h}).  Extending the notation of eq.\ (\ref{ibetamu}) to allow
more than one factor $b_\beta(m)$ in the integrand (e.g.\
$M^{11}_{2}(k,k)$ would have two $b_1(k)$ factors in the integrand),
the 1h-2h term is, to order $1/V$ (there is also a $1/V^2$ term),
\begin{eqnarray}
\expect{\Delta P^{\rm 1h}(k_i)\Delta P^{\rm 2h}(k_j)} =& \frac{1}{V}P^{\rm lin}(k_j)[M^{11}_2(k_i,k_i)M^2_0(k_j,k_j) +\nonumber\\
& 2M^1_1(k_i)M^1_3(k_i,k_j,k_j)].
\label{p1hp2h}
\end{eqnarray}
The second term here (when combined with the same term of
$\expect{\Delta P^{\rm 1h}(k_j)\Delta P^{\rm 2h}(k_i)}$) is the CH
covariance from $T^{\rm 2h}_{31}$; see eq.\ (\ref{t2h31}).  To order
$1/V$, the 2h-2h term is
\begin{eqnarray}
\expect{\Delta P^{\rm 2h}_i\Delta P^{\rm 2h}_j} = & \frac{P^{\rm
lin}(k_i)P^{\rm lin}(k_j)}{V} [M^{11}_2(k_1,k_1)M^1_1(k_2)^2 +\nonumber\\ &
2M^{11}_2(k_1,k_2)M^1_1(k_1)M^1_1(k_2) + \nonumber\\ &
M^{11}_2(k_2,k_2)M^1_1(k_2)^2].  
\label{p2hp2h}
\end{eqnarray}
If $M^{11}_2$ is replaced with $M^2_2$, the $M^{11}$ terms in eqs.\
(\ref{p1hp2h}) and (\ref{p2hp2h}) correspond roughly to $T^{\rm 2h}_{22}$ and
the part of $T^{\rm 3h}$ which does not involve the bispectrum (see CH).

Taken at face value, eqs.\ (\ref{fgint}, \ref{ifigifg}) have
significant implications not only for the power spectrum covariance,
but for the power spectrum itself, since the two-halo term involves
the square of an integral over the mass function (eq.\ \ref{p1h2h}).
By averaging together $P^{\rm 2h}$ from 262144 Poisson-sampled mass
functions, we verified numerically that
\begin{equation}
\expect{\tilde{M}^1_1(k)^2} = M^1_1(k)^2 + \frac{1}{V}M^{11}_2(k,k).
\label{disc2h}
\end{equation}

However, the $1/V$ halo discreteness term cannot contribute to the
two-halo term.  The two-halo term is the convolution of the halo-halo
power spectrum with density profiles of two separate haloes, but in
the halo discreteness calculation, there can only be (zero or) one
halo in each bin, so the `$i=j$' term does not contribute.  A similar
argument applies to terms in the halo-model trispectrum; the only
terms which contain products of integrals over the mass function are
multiple-halo terms.  The covariance from halo discreteness, on the
other hand, is meaningful because it involves a product of complete
power spectrum terms.

\end{document}